\documentclass{aastex62}

\shortauthors{Xu et al.}

\usepackage{amsmath}

\usepackage{CJK}
\usepackage{float}
\usepackage[normalem]{ulem}
\usepackage[shortlabels]{enumitem}
\usepackage{multirow}
  
\def\Gaia  {$Gaia$}

\def\masy    {mas~yr$^{-1}$}

\def\uas     {$\mu$as}

\def\h       {\ifmmode{^{\rm h}}\else{$^{\rm h}$}\fi}
\def\m       {\ifmmode{^{\rm m}}\else{$^{\rm m}$}\fi}
\def\s       {\ifmmode{^{\rm s}}\else{$^{\rm s}$}\fi}
\def\deg     {\ifmmode{^{\circ}}\else{$^{\circ}$}\fi}
\def\decdeg  {\ifmmode{{\rlap.}^{\circ}} \else ${\rlap.}^{\circ}$\fi}
\def\decs    {\ifmmode{{\rlap.}^{\rm s}} \else ${\rlap.}^{\rm s}$\fi}
\def\decas   {\ifmmode{{\rlap.}{''}}\else{${\rlap.}{''}$}\fi}

\def\Vsbar {\ifmmode {\overline{V_s}}\else {$\overline{V_s}$}\fi}
\def\Usbar {\ifmmode {\overline{U_s}}\else {$\overline{U_s}$}\fi}
\def\Wsbar {\ifmmode {\overline{W_s}}\else {$\overline{W_s}$}\fi}

\def\mux    {\ifmmode {\mu_x}\else {$\mu_x$}\fi}
\def\muy    {\ifmmode {\mu_y}\else {$\mu_y$}\fi}
\def\mura   {\ifmmode {\mu_{\alpha}}\else {$\mu_{\alpha}$}\fi}
\def\mude   {\ifmmode {\mu_{\delta}}\else {$\mu_{\delta}$}\fi}

\def\gax{\mathrel{\rlap{\lower4pt\hbox{\hskip1pt$\sim$}}
    \raise1pt\hbox{$>$}}}

\def\d    {\ifmmode {{\rlap{.}}^\circ}\else {${\rlap{.}}^\circ$}\fi}
\def\s    {\ifmmode {{\rlap{.}}^s}\else {${\rlap{.}}^s$}\fi}
\def\as   {\ifmmode {{\rlap{.}}^{''}}\else {${\rlap{.}}^{''}$}\fi}

\begin{document}

\begin{CJK*}{UTF8}{gbsn}

\title{Comparison of {\it Gaia\/} DR2 Parallaxes of Stars with VLBI Astrometry}

\correspondingauthor{Bo Zhang}
\email{zb@shao.ac.cn}

\author[0000-0003-2953-6442]{Shuangjing Xu (徐双敬)}
\affil{Shanghai Astronomical Observatory, Chinese Academy of Sciences, 80 Nandan Road, Shanghai 200030, China}
\affil{University of Chinese Academy of Sciences, 19A Yuquanlu, Beijing 100049, China}

\author[0000-0003-1353-9040]{Bo Zhang (张波)}
\affil{Shanghai Astronomical Observatory, Chinese Academy of Sciences, 80 Nandan Road, Shanghai 200030, China}

\author{Mark J.  Reid}
\affiliation{Center for Astrophysics~$\vert$~Harvard \& Smithsonian, 60 Garden Street, Cambridge, MA 02138, USA}

\author{Xingwu Zheng (郑兴武) }
\affiliation{
School of Astronomy and Space Science, Nanjing University, 22 Hankou Road, Nanjing  210093, China}

\author{Guangli Wang (王广利)}
\affil{Shanghai Astronomical Observatory, Chinese Academy of Sciences, 80 Nandan Road, Shanghai 200030, China}
\affil{University of Chinese Academy of Sciences, 19A Yuquanlu, Beijing 100049, China}

\begin{abstract}
We compare the parallaxes of stars from VLBI astrometry in the literature to those 
in the \Gaia\ DR2 catalog. Our full sample contains young stellar objects, evolved 
AGB stars, pulsars and other radio stars.  
Excluding AGB stars, which show significant discrepancies between \Gaia\ and VLBI 
parallaxes, and stars in binary systems, we obtain an average, systematic, 
parallax offset of  $-75 \pm 29~\mu$as for \Gaia\ DR2, consistent 
with their estimate of a parallax zero-point between $-100$ and 0 $\mu$as.  
\end{abstract}


\section{Introduction}\label{sec:intro}
\label{sec:intro}

The second data release (DR2) of \Gaia\ provides precise celestial
coordinates, trigonometric parallaxes, and proper motions for more than
1.3 billion stars based on observations collected during the first 22
months of the mission (since July 2014) \citep{2018A&A...616A...1G}.  
%
However, there are systematic astrometric errors in \Gaia\
DR2.  The astrometric uncertainties of celestial objects provided in
\Gaia\ DR2 are mainly depended on their magnitudes and celestial
positions owing to the scanning law of \Gaia~\citep{2018A&A...616A...2L}.
For parallaxes, uncertainties are typical $\sim$ 0.04, 0.1 and 0.7 mas
for stars with G magnitude $\le$ $\sim$14, 17 and 20, 
respectively~\citep{2018A&A...616A...9L}.
Similarly to \Gaia\ DR1, all sources are treated as
single stars and thus representable by five astrometric parameters
associated with parallax and proper motion.
For unresolved binaries (separation $\le$ 100 mas), the results thus
refer to the photo-center, and orbital motion and photometric variability
may corrupt the astrometric parameters.  For resolved
binaries, the results may refer to either component and are sometimes
spurious due to confusion of the components.  
Based on quasars and validation solutions, \citet{2018A&A...616A...2L} 
estimate that the zero-point parallax corrections depend on position, 
magnitude, and color and are generally below 100 \uas\ in magnitude, with
an average bias of about $-29$ \uas.  
 
Clearly, independent assessments of $Gaia$ parallaxes are important to fully
characterize systematic errors.
\citet{2018ApJ...862...61S} find a parallax zero-point
of $-82 \pm 33$ \uas\ based on 89 eclipsing binaries.
\citet{2018arXiv180502650Z} present an independent confirmation of
parallax zero-point of $-52.8 \pm 2.4$ \uas\ (stat.)  $\pm 1$ \uas\
(syst.) based on 3500 evolved stars in the $Kepler$ field.
\citet{2018ApJ...861..126R} find the parallax offset to be $-46 \pm 13$
\uas\ based on the $Hubble~Space~Telescope$~($HST$) data of 50
long-period Galactic Cepheids. Based on a direct comparison of the
\Gaia\ DR2 with VLBA parallaxes for 55 young stars,
\citet{2018AJ....156...84K} find a parallax zero-point of $-73 \pm 34$
\uas.    \citet{2018arXiv181207838B} also
estimates a parallax zero-point of $-38 \pm 46$ \uas\ using a sample of 75 radio stars with
\Gaia\ DR2 and  VLBI measurements.
All of these estimated zero-points are larger than the \Gaia\ average value of $-29$ \uas. 
However, the stars from the above-mentioned comparisons are either a
specific stellar class of stars or from a specific region within a
limited distant range and some of the astrometric results are not from
trigonometric parallax measurement. Therefore, a comparison using stars
of different types with independent trigonometric parallaxes could
robustly assess the \Gaia\ DR2 parallax zero-point issue.

Very Long Baseline Interferometry (VLBI) astrometry is capable of
measuring parallaxes with accuracies of $\sim$10 \uas
~\citep{2014ARA&A..52..339R}, which is comparable to or better than the
goals of \Gaia.  Because radio waves are not absorbed significantly by
interstellar dust, the entire Milky Way is available for VLBI observation.
Such measurements have now been carried out for hundreds of radio sources
as distant as 20 kpc~\citep{2017Sci...358..227S}.  Since VLBI astrometry
is relative to distant quasars, the results are absolute parallaxes with
no significant zero-point correction.
Thus, in order to perform an independent
assessment of {\it Gaia\/} DR2 parallaxes directly, we collected a
sample of $\sim$100 stars of different types with published VLBI
parallaxes.  Most of the VLBI parallax uncertainties in the sample are
smaller than those of the \Gaia\ DR2, which offers a unique
opportunity to assess \Gaia\ DR2 parallaxes, particularly for the
\Gaia\ parallaxes zero-point.

\section{Data}\label{sec:data}

\begin{figure}[!ht]
\centering
\includegraphics[width=1.0\linewidth]{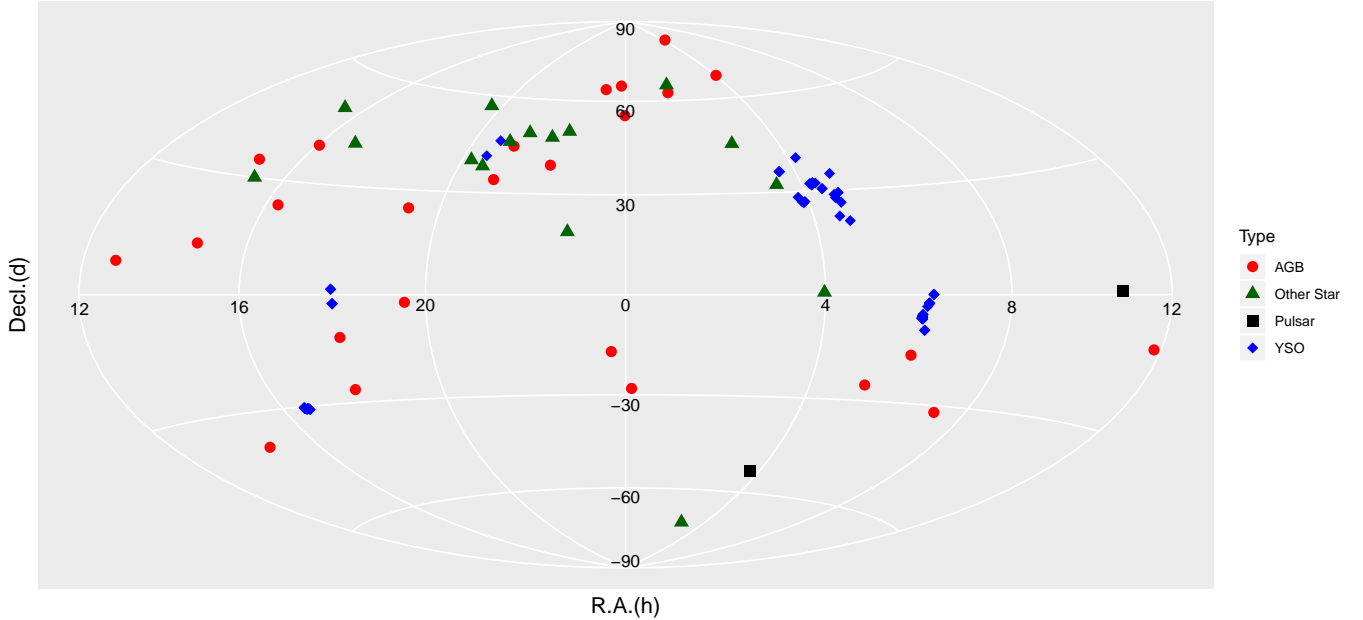}
\caption{The sky distribution of the stars with both VLBI and \Gaia\ DR2
parallaxes. 
}
\label{fig:data} 
\end{figure}

In Table~\ref{tab:data}, we compile a catalog of  108
stars with both VLBI and \Gaia\ DR2 astrometric results. The sky
distribution of these stars are shown in Figure~\ref{fig:data}, with most
of the stars in or near the Galactic plane.
We assembled this catalog by starting with more than 130 stars with 
VLBI parallax results and finding that 93 already have \Gaia\ DR2 results 
be identified in the SIMBAD database \citep{2000A&AS..143....9W}.  
The methods used for the cross-identification of SIMBAD with \Gaia\ DR2 by Simbad Team are as follows:
(1) Objects must have pre-{\it Gaia} sub-arcsecond accurate coordinates in SIMBAD;
(2) Coordinates of {\it Gaia} objects were calculated at epoch 2000.0 taking into account their measured proper motions in order to be compared to the positions in SIMBAD;
(3) {\it Gaia} stars were discarded if they have a neighbouring star in SIMBAD closer than 3\arcsec;
(4) SIMBAD stars were also discarded if they have a neighbouring star at less than 3\arcsec\ in \Gaia\ DR2, 
with a difference of magnitude of $<3$ mag, the brighter source being at less than 1\arcsec\ from the SIMBAD position;
(5) Finally SIMBAD stars that were cross-identified with a \Gaia\ DR2 source with a positional 
difference $>1.0$\arcsec\ were also discarded.
Additionally, we found 15 VLBI/\Gaia\ stars that do not appear in SIMBAD.
Among them there are seven stars (IRAS 18286-0959,
bet Per, GBS-VLA J183123.62-020535.8, 2MASS J16264923-2420029, V1098
Tau, XZ Tau, \& V1000 Tau) with null values for \Gaia\ parallaxes and 
one star (VY CMa) with a negative \Gaia\ parallax, and we do not use them in the comparison.

For VLBI parallaxes, there are 15 stars with more than one parallax measurement,
and for these used variance-weighted averaged results. 
The stellar parallaxes listed in Table~\ref{tab:data}
range from $\approx$ 0.26 to 93.14 mas, corresponding to distances from
$\approx$ 4 to 0.011 kpc.  The median uncertainties of these VLBI and
\Gaia\ DR2 parallaxes are 74 and 123 \uas, respectively.  
The stars in our sample include a number of stellar types, including YSOs, 
AGB stars, pulsars and other radio stars.  

\section{Comparison of VLBI and \Gaia\ DR2 parallaxes }\label{sec:results}

The astrometric results for pulsars in \Gaia\ DR2 are from binary companions 
\citep{2018ApJ...864...26J}.  We find that the parallax result of PSR J1023+0038 
is consistent between VLBI and \Gaia\ DR2, but there is a very large discrepancy 
for PSR J0437-4715.  Since this sub-sample is small and unreliable, owing to
the unmodeled effects of binary orbits on parallax, we do not
consider them further in our analysis.

We examined three independent sub-samples of stars of different stellar types
in order to compare the astrometric results of  VLBI and \Gaia\ DR2.
Our sub-samples are as follows:
\begin{enumerate}
\item AGB: all AGB stars. 
\item YSO: all YSO stars. 
\item ``Other'': the remaining stars in the Table~\ref{tab:data}.
\end{enumerate}
In addition, we consider three groupings of these sub-samples:
\begin{enumerate}[4.]
\item AGB+YSO: the union of the AGB and YSO sub-samples.  
\end{enumerate}
\begin{enumerate}[5.]
\item YSO+''Other'': the union of the YSO and Other sub-samples.
\end{enumerate}
\begin{enumerate}[6.]
\item Full: the union of the AGB, YSO and Other sub-samples. 
\end{enumerate}

\begin{figure}[!ht]
\centering
\includegraphics[width=1.0\linewidth]{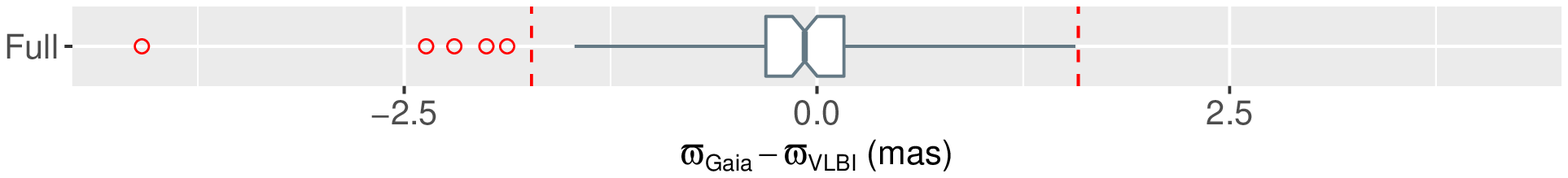}
\includegraphics[width=1.0\linewidth]{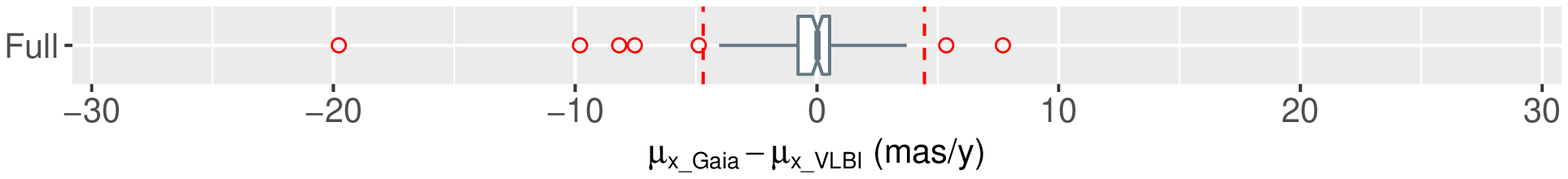}
\includegraphics[width=1.0\linewidth]{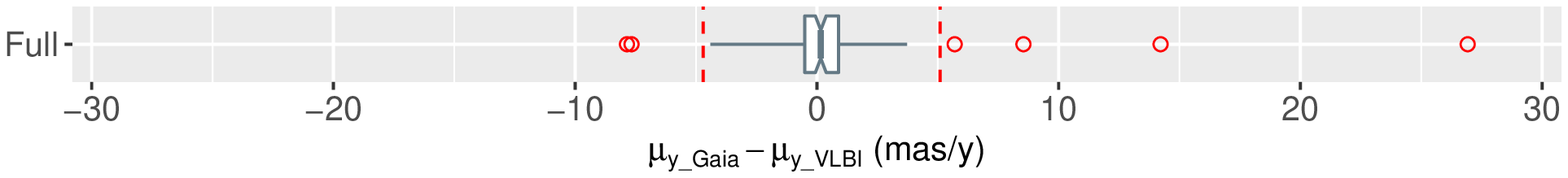}
\caption{Box plots of differences between \Gaia\ DR2 and VLBI for parallax, $\varpi$, 
({\it upper panel}), eastward motion, $\mu_x$, ({\it middle panel}) and northward motion, 
$\mu_y$, ({\it lower panel}).  
The vertical boundaries (``hinges'') indicate the interquartile range (IQR). 
The notches on the sides of the box hinges are at $\pm1.58 \times IQR/\sqrt{N}$, 
representing the standard deviation of the median for a Gaussian distribution. 
Vertical dashed red lines are set at $3.0\times IQR$ above and below the 25\% and 75\% quartiles based on the Full sample.
Extreme outliers fall outside the vertical dashed red lines and were excluded from our analysis. 
}
\label{fig:box} 
\end{figure}

Since we expect some outliers in the \Gaia\ DR2 catalog (eg, from binaries),
we use ``box'' plots \citep{1977eda..book.....T, 2012msma.book.....F}, 
a compact display of robust measures of location and spread, to identify and remove outliers.  
Figure~\ref{fig:box} plots parallax and proper motion differences between 
the DR2 and VLBI measurements for our full sample.  
The boxes denote the inner quartile range ($IQR$), and the vertical red dashed-lines 
at $\pm3.0\times IQR$, separate the extreme outliers from the rest of the data.
There are 5, 7 and 6 extreme outliers in the discrepancies of $\varpi$, $\mu_x$ and $\mu_y$, 
respectively.  These were separately removed from the statistics of each parameter,
so that, for example, a proper motion parameter was still used for a star with a parallax outlier
that was removed from the parallax statistics.

\begin{figure}[!ht]
\centering
\includegraphics[width=1\linewidth]{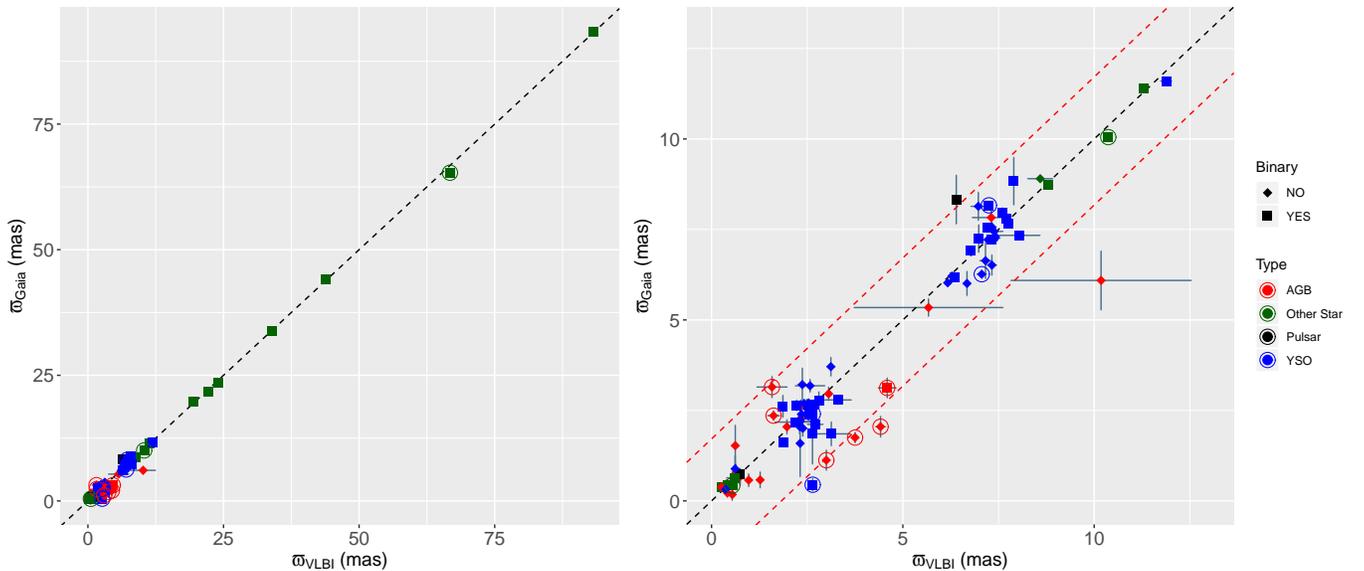}
\caption{\Gaia\ DR2 versus VLBI parallaxes. {\it Left panel:}
all stars in the sample; {\it right panel:} stars with $\varpi<13$ mas.
Colors denote different stellar types and known binaries are shown in different shapes,
as indicated in the legend at the right.
Circles denote differences of $>3\sigma$, where
$\sigma = \sqrt{ {\sigma_{\varpi_{\rm VLBI}}}^2 + {\sigma_{\varpi_{Gaia}}}^2}$.
The black dashed line is a fit with slope of 1 and intercept of 0.
The red dashed lines in right panel separate extreme outliers as shown in Figure~\ref{fig:box}. }
\label{fig:plx}
\end{figure}

\begin{figure}[!ht]
\centering
\includegraphics[width=1\linewidth]{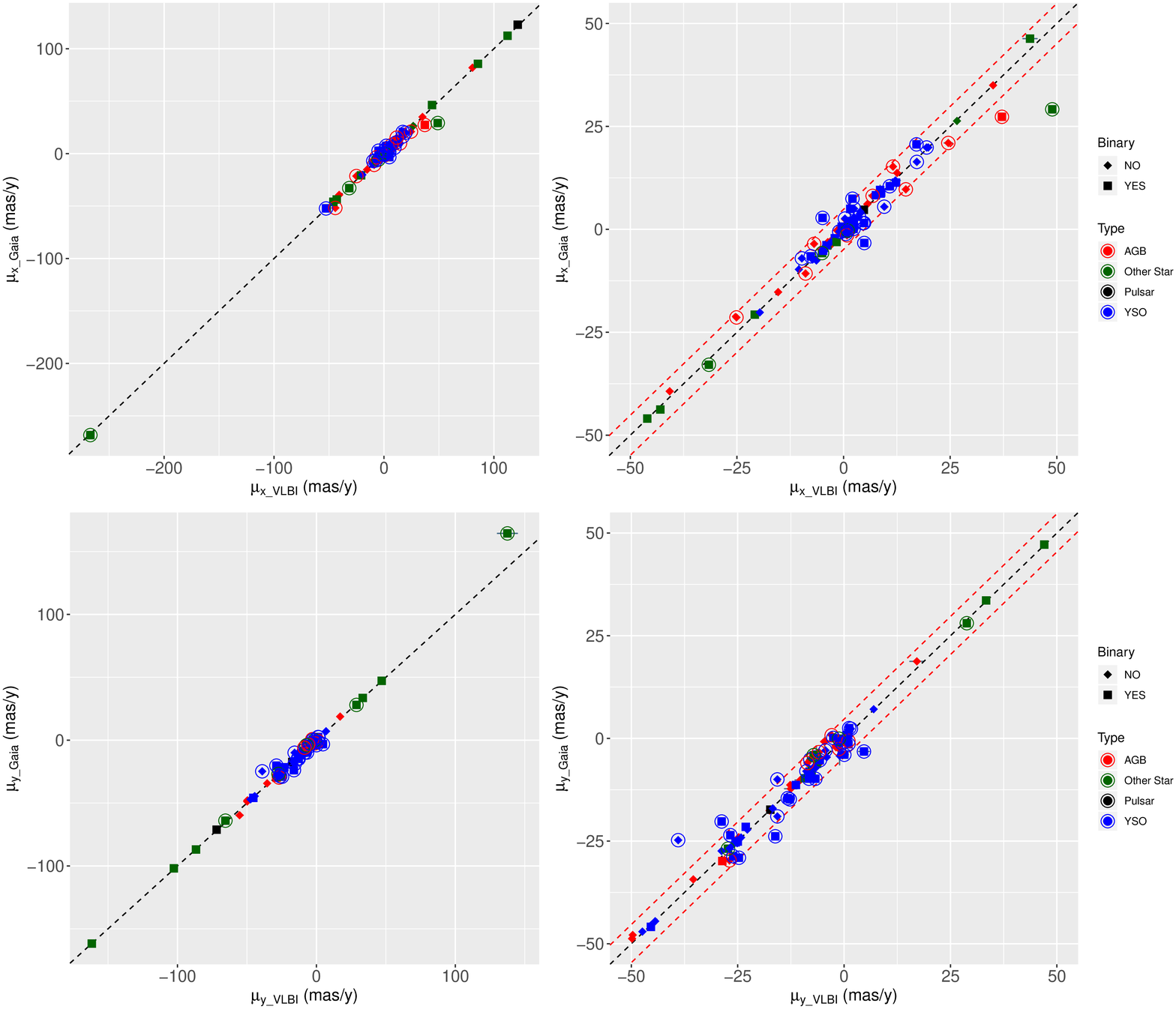}
\caption{\Gaia\ DR2 versus VLBI proper motions. 
{\it Left panels:} all stars in the sample; {\it right panels:} stars with proper motion
components $<50$ mas/yr in magnitude in the eastward {\it top} and northward {\it bottom} directions.
Colors denote different stellar types and binaries are shown in different shapes,
as indicated in the legend at the right.
Circles denote differences of $>3\sigma$, where for each coordinate
$\sigma = \sqrt{ {\sigma_{\mu_{\rm VLBI}}}^2 + {\sigma_{\mu_{\rm Gaia}}}^2}$
The black dashed line is a straight line with slope of 1 and intercept of 0.
The red dashed lines in right panel separate extreme outliers as shown in Figure~\ref{fig:box}. }
\label{fig:pm}
\end{figure}

In order to fit the relationship between the VLBI and \Gaia\ DR2 results,
we adopt linear regression for data with errors in both X and Y axes, following the method 
derived by \citet{Deming1943Statistical} and described in detail in \cite{2004AmJPh..72..367Y}.  
We estimate the slope, intercept, and standard errors of the best straight line
by minimizing the sum:
\begin{eqnarray}
 { \chi  }^{ 2 }=\sum _{ k=1 }^{ n }{ \left[ \frac { { \left( { X }_{ k }-{ x }_{ k } \right)  }^{ 2 } }{ { \sigma  }_{ X,k }^{ 2 } } +\frac { { \left( { Y }_{ k }-{ y }_{ k } \right)  }^{ 2 } }{ { \sigma  }_{ Y,k }^{ 2 } }  \right]  }  
\end{eqnarray}
where (${ { X }_{ k }}$, $ { Y }_{ k }$) denote the $k^{th}$
data pair with corresponding standard deviations 
(${{\sigma}_{X,k}}$, ${{ \sigma  }_{ Y,k }}$) and (${ { x }_{ k }}$, $ { y }_{ k }$) 
denote points of the estimated straight line. 
The best linear fit results are listed in Table~\ref{tab:plx} \& \ref{tab:pm}.  
Figure~\ref{fig:plx} \& \ref{fig:pm} show the direct comparison of the VLBI and \Gaia\ DR2 results.

\begin{deluxetable*}{ccc|cc|c}
\tabletypesize{\scriptsize}
\tablewidth{0pt} 
\tablenum{2}
\tablecaption{\Gaia\ DR2 versus VLBI parallaxes  \label{tab:plx}}
\tablehead{
\multicolumn{1}{c|}{} &
\multicolumn{1}{c|}{}& 
\multicolumn{1}{c|}{} &
\multicolumn{2}{c|}{$\varpi$} &
\multicolumn{1}{c}{ $\varpi$}  \\
\multicolumn{1}{c|}{Sample} & \multicolumn{1}{c|}{Including} &
 \multicolumn{1}{c|}{Number\tablenotemark{a}}& 
\multicolumn{1}{c}{Slope} & \multicolumn{1}{c|}{Intercept} & 
\multicolumn{1}{c}{Weighted mean\tablenotemark{b}} \\
  \multicolumn{1}{c|}{} & \multicolumn{1}{c|}{ Binary?} &
 \multicolumn{1}{c|}{}&  \multicolumn{1}{c}{} & \multicolumn{1}{c|}{(mas)} & 
\multicolumn{1}{c}{(mas)} 
} 
\startdata 
{AGB} & YES    &   19      & $1.008\pm0.080$ &  $-0.001\pm0.105$     & $ +0.006\pm0.078$\\
            &  NO     &  18      &$1.069\pm0.078$ &  $-0.036\pm0.094$      & $ +0.023\pm0.070$    \\
\hline
{YSO} & YES    &   59    &  $0.999\pm0.010$ &  $-0.057\pm0.055$    & $-0.062\pm0.026$  \\
            &  NO    &   33      &   $0.985\pm0.012$ &  $-0.009\pm0.059$   & $-0.076\pm0.029$   \\
\hline
{Other} & YES    &   15    &   $1.004\pm0.003$ &  $-0.103\pm0.050$  &  $-0.075\pm0.047$ \\
            &  NO    &    1    &      &      &    \\
\hline
{AGB+YSO} & YES    &   78    &  $0.996\pm0.009$ &  $-0.037\pm0.046$   &$-0.055\pm0.024$ \\
            &  NO    &   51      &  $0.984\pm0.010$ &  $ +0.003\pm0.047$  &  $-0.061\pm0.027$  \\
\hline
{YSO+Other} & YES    &   74    &   $1.003\pm0.003$ &  $-0.080\pm0.028$    & $-0.065\pm0.022$ \\
            &  NO    &   34      &   $0.986\pm0.011$ &  $-0.011\pm0.058$  & $-0.075\pm0.029$  \\
\hline
{Full} & YES       &  93    & $1.002\pm0.003$ &  $-0.069\pm0.027$     & $-0.059\pm0.022$     \\
            &  NO     &  52    & $0.984\pm0.010$ &  $ +0.001\pm0.047$  & $-0.061\pm0.027$  \\
\hline
\enddata
\tablenotetext{a}{The number of stars with outliers removed in each sample.}
\tablenotetext{b}{Variance weighting as $1/ ({ {\sigma_{\varpi_{\rm VLBI}}}^2 + {\sigma_{\varpi_{Gaia}}}^2})$}
\end{deluxetable*}

\begin{deluxetable*}{cc|cccc|cccc}
\tabletypesize{\scriptsize}
\tablewidth{0pt} 
\tablenum{3}
\tablecaption{Proper motions of \Gaia\ DR2 verses VLBI \label{tab:pm}}
\tablehead{
\multicolumn{1}{c|}{} &
\multicolumn{1}{c|}{}& 
\multicolumn{4}{c|}{$\mu_x$\tablenotemark{c}} &
\multicolumn{4}{c}{$\mu_y$\tablenotemark{c}} \\
\multicolumn{1}{c|}{Sample} & \multicolumn{1}{c|}{Including} &
 \multicolumn{1}{c|}{Number\tablenotemark{a}}& 
\multicolumn{1}{c}{Slope} & \multicolumn{1}{c|}{Intercept} & 
\multicolumn{1}{c|}{Weighted mean\tablenotemark{b}} & 
 \multicolumn{1}{c|}{Number\tablenotemark{a}}& 
\multicolumn{1}{c}{Slope} & \multicolumn{1}{c|}{Intercept} & 
\multicolumn{1}{c}{Weighted mean\tablenotemark{b}} \\
  \multicolumn{1}{c|}{} & \multicolumn{1}{c|}{ Binary?} &
 \multicolumn{1}{c|}{}&  \multicolumn{1}{c}{} & \multicolumn{1}{c|}{(\masy)} & 
\multicolumn{1}{c|}{(\masy)} &
 \multicolumn{1}{c|}{}&  \multicolumn{1}{c}{} & \multicolumn{1}{c|}{(\masy)} & 
\multicolumn{1}{c}{(\masy)} 
} 
\startdata 
{AGB} & YES    &   20      & $1.000\pm0.025$ &  $+0.346\pm0.254$     &$+0.354\pm0.249$ &  23    & $0.995\pm0.032$ &  $+0.430\pm0.396$     &   $+0.464\pm0.318$ \\
            &  NO     &  20      & $1.000\pm0.025$ &  $+0.346\pm0.254$     & $+0.354\pm0.249$ &  22    & $0.985\pm0.034$ &  $+0.398\pm0.397$      &   $+0.503\pm0.325$ \\
\hline
{YSO} & YES    &   57    &   $0.988\pm0.006$ &  $-0.029\pm0.098$    &  $-0.038\pm0.102$ &  55    & $1.008\pm0.011$ &  $-0.014\pm0.213$      &  $-0.130\pm0.141$ \\
            &  NO    &   33    &   $0.985\pm0.011$ &  $+0.071\pm0.124$  &  $-0.001\pm0.114$ &  31    & $0.996\pm0.007$ &  $-0.046\pm0.137$     &  $+0.011\pm0.100$ \\
\hline
{Other} & YES    &   14    &   $1.004\pm0.001$ &  $-0.169\pm0.085$  &  $-0.282\pm0.129$ &  14    & $1.000\pm0.003$ &  $+0.103\pm0.129$     &  $+0.105\pm0.126$ \\
            &  NO    &    1    &      &    &  &    1    &      &        \\
\hline
{AGB+YSO} & YES    &   77    &  $0.988\pm0.006$ &  $-0.002\pm0.088$   &  $-0.010\pm0.091$  &  78    & $1.009\pm0.010$ &  $+0.038\pm0.182$     &  $-0.089\pm0.123$ \\
            &  NO    &   53      &  $0.985\pm0.010$ &  $+0.103\pm0.104$ & $+0.042\pm0.100$   &  53    & $0.998\pm0.007$ &  $+0.047\pm0.142$     &  $+0.078\pm0.105$ \\
\hline
{YSO+Other} & YES    &   71    &  $1.003\pm0.001$ &  $-0.091\pm0.079$   &  $-0.126\pm0.081$ &  69    & $1.002\pm0.005$ &  $-0.049\pm0.133$    &  $-0.081\pm0.115$ \\
            &  NO    &   34      &   $0.985\pm0.011$ &  $+0.071\pm0.122$  &  $-0.003\pm0.112$  &  32    & $0.996\pm0.006$ &  $-0.039\pm0.134$    &  $+0.014\pm0.098$ \\
\hline
{Full} & YES    &  91    & $1.003\pm0.001$ &  $-0.069\pm0.072$      &  $-0.103\pm0.075$   &  92    & $1.003\pm0.005$ &  $-0.016\pm0.120$     &  $-0.051\pm0.104$ \\
            &  NO     &  54    & $0.985\pm0.009$ &  $+0.103\pm0.103$   &  $+0.040\pm0.099$ &  54   &  $0.998\pm0.007$ &  $+0.051\pm0.141$    &   $+0.081\pm0.104$ \\
\hline
\enddata
\tablenotetext{a}{The number of stars with outliers removed in each sample.}
\tablenotetext{b}{Variance weighting as $1/ ( {\sigma_{\mu_x{\rm VLBI}}}^2 + {\sigma_{\mu_x{\rm Gaia}}}^2)$ and  $1/ ( {\sigma_{\mu_y{\rm VLBI}}}^2 + {\sigma_{\mu_y{\rm Gaia}}}^2)$} 
\tablenotetext{c}{$\mu_x=\mu_{\alpha} \cos{\delta}$, $\mu_y=\mu_{\delta}$}
\end{deluxetable*}

As shown in Table~\ref{tab:plx}, the agreement between the \Gaia\ DR2 and VLBI parallaxes
is excellent for all sub-samples, with the slopes of the fitted lines consistent with unity 
within $2\sigma$ uncertainties.  The (y-axis) intercepts are small and negative, indicating
the \Gaia\ parallaxes are systematically smaller than the (absolute) VLBI parallaxes,
as expected for the \Gaia\ DR2 catalog.
 
\begin{figure}[!ht]
   \centering
         \includegraphics[width=0.6\linewidth]{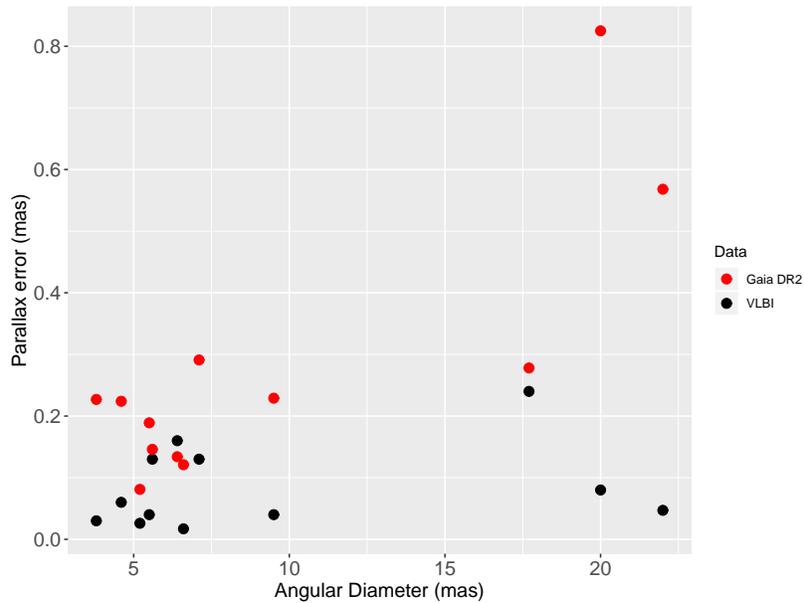}
   \caption{Parallax uncertainties versus angular diameters for AGB stars. {\it Red} and 
{\it black dots} denote \Gaia\ DR2 and VLBI parallaxes, respectively.}
    \label{fig:size}
\end{figure}

Among all the samples, the AGB sample has the largest uncertainties and discrepancies
between \Gaia\ DR2 and VLBI data. Stellar angular diameters for 12 AGB stars from literature 
are listed in Table ~\ref{tab:size}.  In Figure~\ref{fig:size}, we plot the \Gaia\ DR2 and 
VLBI parallax uncertainties for these AGB stars versus their angular diameters. 
There is a clear trend that the larger of the stellar size, the larger of the stellar parallax 
uncertainty for \Gaia\ DR2, while there is no such a trend for VLBI observations.  
This is not unexpected for stars with angular sizes comparable to their parallaxes and
known to have significant surface brightness variations.

\begin{figure}[!ht]
   \centering
            \includegraphics[width=1\linewidth]{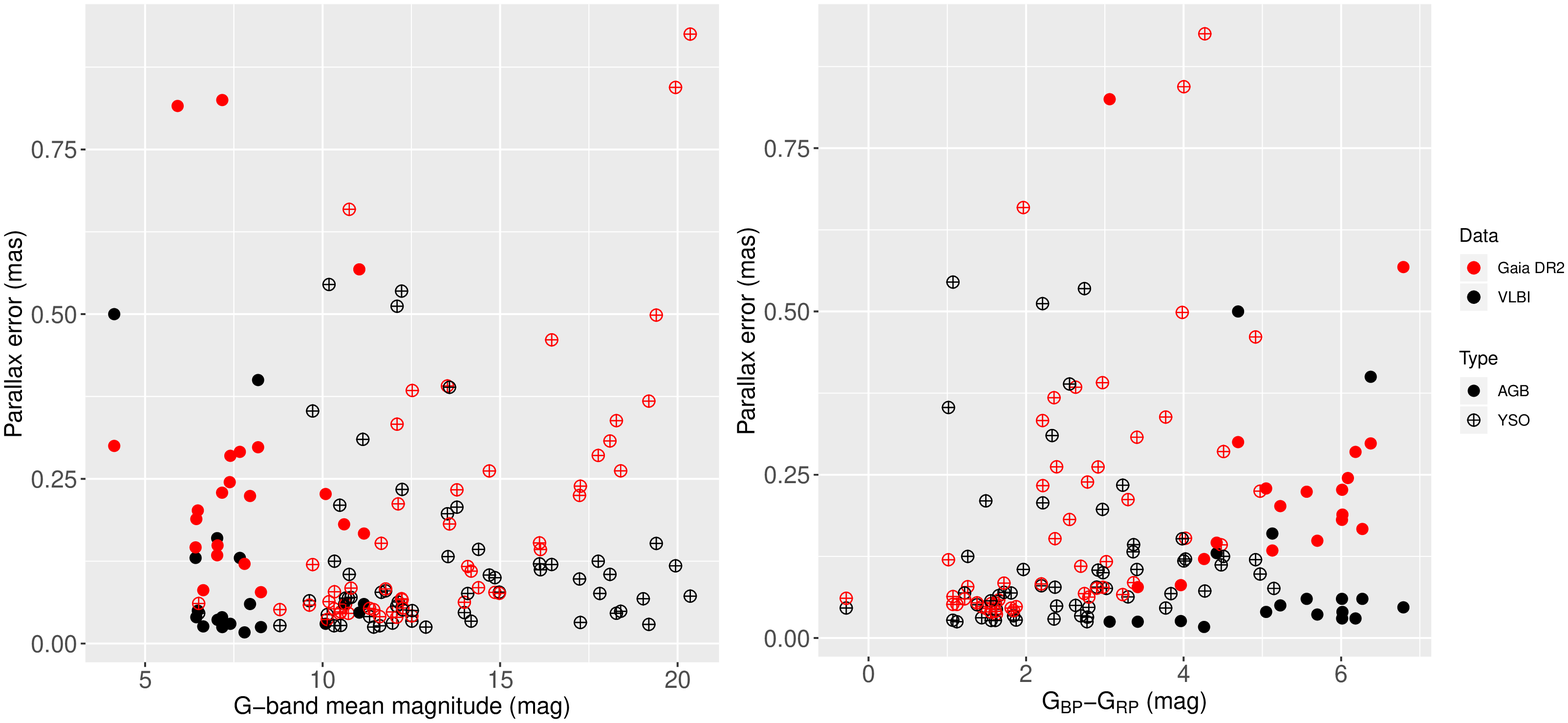}
   \caption{
{\it Left panel} Parallax uncertainty versus G-band mean magnitude.  
{\it Right panel} Parallax uncertainty versus color ($G_{BP}$-$G_{RP}$). 
$G_{BP}$ and $G_{RP}$ are the \Gaia\ magnitudes of the blue and red photometer, respectively.}
    \label{fig:mag}
\end{figure}

As mentioned by \citet{2018A&A...616A...9L}, the systematic errors in \Gaia\ DR2 are complicated, 
affected by position on the sky, magnitude, and color. \citet{2018A&A...618A..58M} claimed that 
the precision reached on the parallax depends on the $G_{BP}$-$G_{RP}$ color for long period 
variable candidates, where the $G_{BP}$ and $G_{RP}$ are the \Gaia\ magnitudes of blue and red 
photometer values, respectively.  The $G_{BP}$-$G_{RP}$ spread originates from extinction due to 
interstellar and/or circumstellar dust \citep{2018A&A...618A..58M}.  
As shown in Figure~\ref{fig:mag},  the parallax uncertainties of \Gaia\ DR2 are larger than VLBI 
at magnitude $>$ 16 and large $G_{BP}$-$G_{RP}$. This shows that redder AGB stars give larger 
parallax uncertainties; possible reasons include 1) they tend to be larger, 
2) they probably have more surface brightness variations, and 3) they will likely have more 
circumstellar dust to corrupt their images. We conclude that one should be cautious when using 
the \Gaia\ parallaxes for AGB stars.
 
Certainly, binaries can yield unreliable parallaxes, since unmodeled orbital motions
can be significant.
Note that the sub-samples excluding binaries have larger slope uncertainties
than samples including binaries.  This is simply a result of much larger parallax range
of $\sim93$ mas for samples including binaries than $\sim8$ mas when the binaries are excluded
as shown in  Figure~\ref{fig:plx}.

For our most reliable estimate of the \Gaia\ DR2 parallax zero-point we choose the results of the  
YSO+Other samples that exclude binaries and red giants.
Assuming that the slopes are exactly unity, we can simply calculate a variance-weighted mean 
difference between DR2 and VLBI parallaxes in order to estimate the \Gaia\ zero-point 
correction.  These are also shown in Table~\ref{tab:plx}.
We find the zero-point is $-75\pm29$ \uas\ if the slope is assumed to be exactly unity.
The magnitude of our estimated parallax zero-point is statistically consistent, but perhaps larger 
than that of the \Gaia\ DR2 claim of an overall parallax zero-point of $-29$ \uas.  

As shown in Table~\ref{tab:pm} and Figure~\ref{fig:pm},  the fractional uncertainties of proper 
motions differences, while excellent, is not as good as for the parallaxes.
Our results are independent of those estimated by \citet{2018ApJ...862...61S}, \citet{2018arXiv180502650Z} and \citet{2018ApJ...861..126R}  as mentioned in \S~\ref{sec:intro},
since the stars and the methods for deriving parallaxes are different.
In contrast to  the results of the \citet{2018arXiv181207838B} and \citet{2018AJ....156...84K},  
we use a larger sample of stars with VLBI parallaxes and consider the problems of AGB stars and binaries.
There are  81 overlapping stars in Bobylev's sample (88 stars) and our sample (108 stars);  
the remaining 7 stars in Bobylev's sample had no VLBI parallax results, but were in the sample
as they had proper motions. 
Also, Bobylev did not perform a rigorous least-squares analysis that takes into account errors in both axes.   
Finally, \citet{2018arXiv181207838B}  calculated the weighted mean with weights inversely proportional to the measurement errors $1/ \sqrt { {\sigma_{\varpi_{\rm VLBI}}}^2 + {\sigma_{\varpi_{Gaia}}}^2}$, whereas we use the variance weighting $1/ ({ {\sigma_{\varpi_{\rm VLBI}}}^2 + {\sigma_{\varpi_{Gaia}}}^2})$.
Using our methods on Bobolev's sample, we estimate the parallax zero-point as $-65 \pm 34~\mu$as with 28 stars 
(i.e., excluding AGB and binary stars), which is consistent with our result.

\begin{deluxetable*}{c rrrrrrrrrrrrr}
\tiny
\tablecaption{ Angular diameters AGB stars \label{tab:size}}
\tablenum{4}
\tablewidth{0.001pt}
\tablehead{
  \colhead{Num} &
  \colhead{SIMBAD} &
  \colhead{Angular Diameter} &
  \colhead{Reference} \\
    \colhead{} &
  \colhead{Name} &
  \colhead{(mas)} &
  \colhead{} 
  }
\startdata
1&  S Crt	   & 5.6 $\pm$ 1	  & \citet{2008PASJ...60.1013N} \\
2&  S Per	   & 6.6 $\pm$ 2.5  & \citet{2012AA...546A..16R} \\
3&  SY Scl	 & 3.8  	        & \citet{2011PASJ...63...63N} \\
4& NML Cyg	   & 22	            & \citet{2012AA...544A..42Z} \\
5&  VY CMa	 & 20             & \citet{2000ApJ...543..868M} \\	
6&  PZ Cas	 & 5.2 $\pm$ 1.2	& \citet{2005ApJ...628..973L} \\	
7&  R Aqr	   & 17.7    	      & \citet{2005ApJ...620..961M} \\
8&  T Lep	   & 5.5 $\pm$ 2.4	& \citet{2009AA...496L...1L} \\
9&  RW Lep	 & 6.4  		      & \citet{2014PASJ...66..107K} \\
10&  U Lyn	 & 4.6 $\pm$ 0.4	& \citet{2016PASJ...68...71K} \\	
11&  RT Vir	 & 7.1 $\pm$ 0.4	& \citet{2012AA...546A..16R} \\
12&  VX Sgr	 & 9.5 $\pm$ 0.9	& \citet{2012AA...546A..16R} \\
\enddata
\tablecomments{Some angular diameters are derived from stellar radii and VLBI distances.}
\end{deluxetable*}

\section{Summary}

Based on the comparison of stellar parallaxes from 
\Gaia\ DR2 and the generally more accurate VLBI values, we find a \Gaia\ parallax zero-point 
correction of  $-75 \pm 29~\mu$as.  This value is consistent with published \Gaia\ estimates that
the zero-point is negative and below 100 \uas\ in magnitude.  We find that AGB stars have the 
largest discrepancies in both parallax and proper motions of the samples we consider.   
This is reasonable since AGB stars are generally large, variable, and often surrounded by copious dust.  
Radio observations of circumstellar masers for AGB stars can yield parallax accuracy 
roughly an order-of-magnitude better than \Gaia\ DR2 \citep{2017ApJ...849...99Z,2018ApJ...859...14X}.
In the future, with improvement in sensitivity and calibration of VLBI arrays, it 
should be feasible to detect weaker radio stars and provide a better 
assessment of \Gaia\ parallax accuracy.

\acknowledgments
This work was partly supported by the 100 Talents Project of the
Chinese Academy of Sciences, the National Science Foundation of China
under grant 11673051, 11873077 and U1831136, and the Key Laboratory for Radio
Astronomy, Chinese Academy of Sciences.

\vspace{5mm}
\facilities{\Gaia, VLBA, VERA, EVN, LBA}

\software{
          R~\citep{2013R.Manual.RCoreT},
          TOPCAT~\citep{2005ASPC..347...29T} .
          }

\begin{longrotatetable}
\begin{deluxetable*}{c rrrrrrrrrrrrr}
\tiny
\tablecaption{Parallaxes of \Gaia\ DR2 and VLBI \label{tab:data}}
\tablenum{1}
\tablewidth{0.001pt}
\tabletypesize{\tiny}
\tablehead{
  \colhead{Num} &
  \colhead{Main name in SIMBAD} &
  \colhead{Binary} &
  \colhead{Type} &
  \colhead{VLBI $\varpi$} &
  \colhead{VLBI  $\mu_x$} &
  \colhead{VLBI  $\mu_y$} &
  \colhead{Ref.\tablenotemark{a}} &
  \colhead{\Gaia\ $\varpi$} &
  \colhead{\Gaia\ $\mu_x$\tablenotemark{b}} &
  \colhead{\Gaia\ $\mu_y$\tablenotemark{b}} &
  \colhead{Noise\_sig\tablenotemark{c}} &
  \colhead{Gmag \tablenotemark{d}} &
  \colhead{bp\_rp\tablenotemark{e}}  \\
    \colhead{} &
  \colhead{} &
  \colhead{} &
  \colhead{} &
  \colhead{(mas)} &
  \colhead{(\masy)} &
  \colhead{(\masy)} &
  \colhead{} &
  \colhead{(mas)} &
  \colhead{(\masy)} &
  \colhead{(\masy)} &
  \colhead{} &
  \colhead{(mag)} &
  \colhead{(mag)} 
  }
\startdata
 1   &  R Cas                     &        & AGB        & 5.67$\pm$1.95    & 80.52$\pm$2.35    & 17.1$\pm$1.75     &  5 & 5.342$\pm$0.245  & 81.92$\pm$0.403    & 18.76$\pm$0.358     & 1.77   & 7.383  & 6.086\\
  2   &  W Hya                     &        & AGB        & 10.18$\pm$2.36   & -44.24$\pm$2.04   & -55.28$\pm$2.93   &  5 & 6.091$\pm$0.816  & -51.773$\pm$1.298  & -59.688$\pm$1.256   & 3.327  & 5.915  & \\
  3   &  S CrB                     &        & AGB        & 2.31$\pm$0.33    & -9.08$\pm$0.27    & -12.49$\pm$0.33   &  5 & 2.322$\pm$0.285  & -10.731$\pm$0.473  & -11.348$\pm$0.366   & 0.882  & 7.4    & 6.183\\
  3   &                              &        &            & 2.39$\pm$0.17    & -8.58$\pm$0.38    & -13.21$\pm$0.61   &  8 &                  &                    &                     &        &        & \\
  4   &  U Her                     &        & AGB        & 3.61$\pm$1.04    & -14.94$\pm$0.38   & -9.17$\pm$0.42    &  5 & 1.749$\pm$0.149  & -15.241$\pm$0.213  & -10.141$\pm$0.227   & 0.696  & 7.04   & 5.698\\
  4   &                              &        &            & 3.76$\pm$0.27    & -16.99$\pm$0.77   & -11.88$\pm$0.5    &  8 &                  &                    &                     &        &        & \\
  5   &  RR Aql                    &        & AGB        & 1.58$\pm$0.4     & -25.11$\pm$0.74   & -49.82$\pm$0.54   &  8 & 3.146$\pm$0.298  & -21.397$\pm$0.481  & -48.743$\pm$0.293   & 1.001  & 8.182  & 6.376\\
  6   &  S Crt                     &        & AGB        & 2.33$\pm$0.13    & -3.17$\pm$0.22    & -5.41$\pm$0.22    & 12 & 2.646$\pm$0.146  & -4.039$\pm$0.242   & -4.95$\pm$0.153     & 0.539  & 6.421  & 4.42\\
  7   &  R Aqr                     & Binary\tablenotemark{f} & AGB        & 4.59$\pm$0.24    & 37.13$\pm$0.47    & -28.62$\pm$0.44   & 32 & 3.122$\pm$0.278  & 27.33$\pm$0.423    & -29.859$\pm$0.397   & 0.947  & 7.032  & 5.803\\
  8   &  SY Scl                    &        & AGB        & 0.75$\pm$0.03    & 5.57$\pm$0.04     & -7.32$\pm$0.12    & 21 & 0.675$\pm$0.227  & 6.111$\pm$0.326    & -7.475$\pm$0.29     & 0.728  & 10.085 & 6.013\\
  9   &  RX Boo                    &        & AGB        & 7.31$\pm$0.5     & 24.55$\pm$1.06    & -49.67$\pm$2.38   & 26 & 7.829$\pm$0.3    & 20.978$\pm$0.515   & -47.861$\pm$0.5     & 1.651  & 4.129  & 4.692\\
  10  &  T Lep                     &        & AGB        & 3.06$\pm$0.04    & 14.6$\pm$0.5      & -35.43$\pm$0.79   & 33 & 2.959$\pm$0.189  & 9.713$\pm$0.24     & -34.322$\pm$0.238   & 0.771  & 6.442  & 6.017\\
  11  &  RW Lep                    &        & AGB        & 1.62$\pm$0.16    & 12.55$\pm$0.59    & -26.92$\pm$0.68   & 34 & 2.355$\pm$0.134  & 13.689$\pm$0.231   & -29.712$\pm$0.25    & 0.548  & 7.029  & 5.127\\
  12  &  U Lyn                     &        & AGB        & 1.27$\pm$0.06    & 0.8$\pm$0.57      & -6$\pm$0.56       & 37 & 0.58$\pm$0.224   & -1.457$\pm$0.208   & -6.297$\pm$0.22     & 0.38   & 7.955  & 5.563\\
  13  &  R UMa                     &        & AGB        & 1.97$\pm$0.05    & -40.77$\pm$0.39   & -24.75$\pm$0.38   & 39 & 2.045$\pm$0.202  & -39.334$\pm$0.389  & -24.059$\pm$0.351   & 0.559  & 6.484  & 5.228\\
  14  &  RT Vir                    &        & AGB        & 4.417$\pm$0.13   & 35.056$\pm$0.697  & -17.5$\pm$0.7     & 43 & 2.05$\pm$0.291   & 34.982$\pm$0.624   & -17.552$\pm$0.612   & 1.509  & 7.668  & \\
  15  &  VX Sgr                    &        & AGB        & 0.64$\pm$0.04    & 0.36$\pm$0.76     & -2.92$\pm$0.78    & 45 & 0.787$\pm$0.229  & 2.451$\pm$0.45     & 0.771$\pm$0.397     & 0.907  & 7.167  & 5.048\\
  16  &  VY CMa                    &        & AGB        & 0.83$\pm$0.08    & -2.8$\pm$0.2      & 2.6$\pm$0.2       & 23 & -5.917$\pm$0.825 & 0.926$\pm$1.772    & -6.474$\pm$1.754    & 4.484  & 7.173  & 3.061\\
  16  &                              &        &            & 0.88$\pm$0.08    & -1.882$\pm$0.144  & 1.02$\pm$0.61     & 11 &                  &                    &                     &        &        & \\
  17  & NML Cyg                      &        & AGB        & 0.62$\pm$0.047   & -1.55$\pm$0.42    & -4.59$\pm$0.41    & 22 & 1.526$\pm$0.568  & -0.268$\pm$1.124   & -0.863$\pm$1.179    & 3.952  & 11.033 & 6.791\\
  18  &  S Per                     &        & AGB        & 0.413$\pm$0.017  & -0.49$\pm$0.23    & -1.19$\pm$0.2     & 17 & 0.222$\pm$0.121  & -0.01$\pm$0.295    & -2.57$\pm$0.307     & 0.446  & 7.8    & 4.26\\
  19  &  UX Cyg                    &        & AGB        & 0.54$\pm$0.06    & -6.91$\pm$0.75    & -12.52$\pm$1.57   &  7 & 0.176$\pm$0.167  & -3.529$\pm$0.307   & -12.266$\pm$0.402   & 0.586  & 11.164 & 6.27\\
  20  &  FV Boo                    &        & AGB        & 0.97$\pm$0.06    & 6.81$\pm$0.14     & 1.01$\pm$0.12     & 38 & 0.573$\pm$0.181  & 8.121$\pm$0.319    & -0.602$\pm$0.298    & 0.644  & 10.605 & 6.011\\
  21  &  PZ Cas                    &        & AGB        & 0.356$\pm$0.026  & -3.7$\pm$0.2      & -2$\pm$0.3        & 28 & 0.42$\pm$0.081   & -3.11$\pm$0.118    & -1.808$\pm$0.11     & 0.381  & 6.637  & 3.965\\
  22  & IRAS 18286-0959              &        & AGB        & 0.277$\pm$0.041  & -3.152$\pm$0.295  & -7.2$\pm$0.2      & 30 &                  &                    &                     & 0.0    & 19.97  & 2.035\\
  23  & IRC +60370                   &        & AGB        & 0.4$\pm$0.025    & -1.278$\pm$0.164  & -1.91$\pm$0.17    & 27 & 0.479$\pm$0.078  & -2.655$\pm$0.128   & -2.16$\pm$0.126     & 0.324  & 8.262  & 3.418\\
  24  & HD 283447                    & Binary\tablenotemark{f} & YSO        & 7.692$\pm$0.085  & 10.253$\pm$0.843  & -25.119$\pm$0.301 & 46 & 7.805$\pm$0.141  & 8.932$\pm$0.39     & -29.054$\pm$0.25    & 0.514  & 9.984  & 1.795\\
  24  &                                                &        &            & 7.70$\pm$0.19  & 8.3$\pm$0.50  & -23.6$\pm$0.5 & 49 &                  &                    &                     &        &        & \\
  25  &  V1271 Tau                 &        & YSO        & 7.418$\pm$0.025  & 22$\pm$2          & -45.7$\pm$2.1     & 35 & 7.307$\pm$0.052  & 19.88$\pm$0.111    & -45.515$\pm$0.063   & 0.0    & 11.438 & 1.122\\
  26  &  V811 Tau                  &        & YSO        & 7.223$\pm$0.057  & 20$\pm$2          & -47.9$\pm$6.9     & 35 & 7.215$\pm$0.04   & 19.879$\pm$0.077   & -44.511$\pm$0.05    & 0.0    & 12.087 & 1.552\\
  27  &  V1065 Tau                 &        & YSO        & 7.382$\pm$0.031  & 17.3$\pm$0.7      & -44.8$\pm$1.8     & 35 & 7.221$\pm$0.048  & 16.38$\pm$0.084    & -47.044$\pm$0.054   & 0.0    & 11.968 & 1.437\\
  28  &  V1282 Tau                 & Binary\tablenotemark{f} & YSO        & 7.324$\pm$0.044  & 17.1$\pm$1        & -45.4$\pm$0.7     & 35 & 7.209$\pm$0.051  & 20.641$\pm$0.1     & -45.881$\pm$0.059   & 0.0    & 10.421 & 1.041\\
  29  & [SVS76] Ser 14               &        & YSO        & 2.313$\pm$0.078  & 3.634$\pm$0.05    & -8.864$\pm$0.127  & 42 & 2.241$\pm$0.076  & 2.98$\pm$0.13      & -8.002$\pm$0.113    & 0.431  & 14.977 & 2.896\\
  30  & 2MASS J18300065+0113402      &     Binary\tablenotemark{g}     & YSO        & 2.638$\pm$0.118  & 1.573$\pm$0.07    & -6.513$\pm$0.152  & 42 & 1.858$\pm$0.844  & 4.97$\pm$1.872     & -3.939$\pm$2.08     & 3.556  & 19.942 & 4.003\\
  31  & NAME W 40 IRS 5              &        & YSO        & 2.302$\pm$0.063  & 0.186$\pm$0.053   & -6.726$\pm$0.121  & 42 & 2.053$\pm$0.212  & 0.546$\pm$0.401    & -7.213$\pm$0.339    & 0.787  & 12.128 & 3.294\\
  32  & GBS-VLA J183123.62-020535.8  &        & YSO        & 2.186$\pm$0.076  & -0.258$\pm$0.058  & -7.514$\pm$0.135  & 42 &                  &                    &                     & 4.644  & 17.808 & 5.146\\
  33  & 2MASS J18312601-0205169      &        & YSO        & 2.372$\pm$0.12   & 4.586$\pm$0.074   & -7.946$\pm$0.167  & 42 & 3.212$\pm$0.461  & 1.484$\pm$0.889    & -5.098$\pm$0.733    & 3.0    & 16.452 & 4.914\\
  34  & 2MASS J18312745-0205118      &        & YSO        & 2.385$\pm$0.098  & -0.33$\pm$0.049   & -7.746$\pm$0.111  & 42 & 2.006$\pm$0.225  & -1.507$\pm$0.469   & -8.627$\pm$0.508    & 1.275  & 17.236 & 4.971\\
  35  & 2MASS J16255609-2430148      &        & YSO        & 7.33$\pm$0.112   & -9.78$\pm$0.09    & -25.11$\pm$0.2    & 40 & 7.561$\pm$0.143  & -7.055$\pm$0.35    & -25.498$\pm$0.225   & 0.661  & 16.135 & 4.478\\
  36  & 2MASS J16255752-2430317      &        & YSO        & 7.404$\pm$0.143  & -7.26$\pm$0.04    & -25.29$\pm$0.07   & 40 & 7.261$\pm$0.085  & -7.343$\pm$0.209   & -25.022$\pm$0.13    & 0.41   & 14.392 & 3.365\\
  37  & 2MASS J16263416-2423282      & Binary\tablenotemark{f} & YSO        & 7.249$\pm$0.091  & -2.05$\pm$0.02    & -26.72$\pm$0.04   & 40 & 8.166$\pm$0.112  & -2.17$\pm$0.253    & -23.557$\pm$0.162   & 0.65   & 14.137 & 4.208\\
  38  & CoKu HP Tau G2               &        & YSO        & 6.145$\pm$0.029  & 11.248$\pm$0.022  & -15.686$\pm$0.013 & 46 & 6.026$\pm$0.048  & 11.873$\pm$0.138   & -9.985$\pm$0.08     & 0.0    & 10.51  & 1.874\\
  38  &                              &        &            & 6.2$\pm$0.03     & 13.85$\pm$0.03    & -15.4$\pm$0.2     & 14 &                  &                    &                     &        &        & \\
  39  & Haro 1-6                     &        & YSO        & 7.385$\pm$0.234  & -19.63$\pm$0.19   & -26.92$\pm$0.13   & 40 & 7.446$\pm$0.067  & -20.184$\pm$0.191  & -26.765$\pm$0.119   & 0.178  & 12.238 & 3.227\\
  40  & 2MASS J16264375-2416333      &        & YSO        & 7.16$\pm$0.152   & -10.48$\pm$0.16   & -38.99$\pm$0.35   & 40 & 6.637$\pm$0.499  & -9.741$\pm$1.106   & -24.773$\pm$0.693   & 1.886  & 19.399 & 3.983\\
  41  & 2MASS J16264923-2420029      &        & YSO        & 7.232$\pm$0.068  & -11.62$\pm$0.06   & -18.3$\pm$0.15    & 40 &                  &                    &                     & 9.28   & 19.031 & 3.839\\
  42  & 2MASS J16273084-2447268      &        & YSO        & 7.327$\pm$0.125  & -4.41$\pm$0.11    & -28.79$\pm$0.33   & 40 & 6.515$\pm$0.286  & -5.209$\pm$0.63    & -27.414$\pm$0.408   & 1.513  & 17.764 & 4.507\\
  43  & 2MASS J16303563-2434188      &      Binary\tablenotemark{g}    & YSO        & 7.206$\pm$0.08   & -7.69$\pm$0.02    & -26.04$\pm$0.04   & 40 & 7.551$\pm$0.083  & -6.547$\pm$0.141   & -28.693$\pm$0.104   & 0.12   & 11.774 & 2.193\\
  44  & 2MASS J16315211-2456156      &        & YSO        & 6.676$\pm$0.046  & -6.38$\pm$0.03    & -22.74$\pm$0.04   & 40 & 6.007$\pm$0.338  & -7.568$\pm$0.567   & -22.208$\pm$0.426   & 1.072  & 18.274 & 3.773\\
  45  & DoAr 51                      &     Binary\tablenotemark{g}     & YSO        & 6.983$\pm$0.05   & -4.8$\pm$0.08     & -23.11$\pm$0.11   & 40 & 7.248$\pm$0.384  & -5.196$\pm$1.068   & -21.538$\pm$0.718   & 1.367  & 12.524 & 2.629\\
  46  &  V1096 Tau                 &     Binary\tablenotemark{g}     & YSO        & 8.055$\pm$0.535  & 2.089$\pm$0.73    & -16.167$\pm$0.711 & 46 & 7.337$\pm$0.068  & 7.44$\pm$0.156     & -23.835$\pm$0.105   & 0.167  & 12.223 & 2.741\\
  46  &                              &        &            & 7.924$\pm$1.334  & 7.14$\pm$2.149    & -28.765$\pm$2.186 & 46 &                  &                    &                     &        &        & \\
  47  &  V1098 Tau                 &        & YSO        & 8.07$\pm$0.31    & 11.148$\pm$0.175  & -27.327$\pm$0.172 & 46 &                  &                    &                     & 16.17  & 11.132 & 2.33\\
  48  &  V410 Tau                  &     Binary\tablenotemark{g}     & YSO        & 7.751$\pm$0.027  & 8.703$\pm$0.017   & -24.985$\pm$0.02  & 46 & 7.667$\pm$0.053  & 8.683$\pm$0.128    & -25.1$\pm$0.061     & 0.0    & 10.321 & 1.557\\
  49  &  V1023 Tau                 &     Binary\tablenotemark{g}     & YSO        & 7.686$\pm$0.032  & 8.371$\pm$0.02    & -25.49$\pm$0.02   & 46 & 7.951$\pm$0.152  & 8.296$\pm$0.356    & -25.309$\pm$0.229   & 0.58   & 11.652 & 2.367\\
  49  &                              &        &            & 7.53$\pm$0.03    & 3.785$\pm$0.044   & -28.9$\pm$0.3     &  9 &                  &                    &                     &        &        & \\
  50  &  V1201 Tau                 &     Binary\tablenotemark{g}     & YSO        & 6.363$\pm$0.069  & 10.839$\pm$0.05   & -13.235$\pm$0.058 & 46 & 6.166$\pm$0.046  & 10.469$\pm$0.104   & -14.544$\pm$0.077   & 0.0    & 10.725 & 1.81\\
  51  & HD 283641                    &        & YSO        & 6.285$\pm$0.07   & 10.913$\pm$0.037  & -16.772$\pm$0.044 & 46 & 6.212$\pm$0.084  & 10.612$\pm$0.173   & -17.047$\pm$0.134   & 0.2    & 10.799 & 1.719\\
  52  &  XZ Tau                    &     Binary\tablenotemark{g}     & YSO        & 6.793$\pm$0.025  & 10.858$\pm$0.027  & -16.264$\pm$0.06  & 46 &                  &                    &                     & 5.03   & 12.908 & 2.773\\
  53  &  V807 Tau                  &   Binary\tablenotemark{g}       & YSO        & 7.899$\pm$0.105  & 8.573$\pm$0.068   & -28.774$\pm$0.201 & 46 & 8.834$\pm$0.659  & 9.559$\pm$1.235    & -20.23$\pm$0.989    & 2.994  & 10.749 & 1.965\\
  54  &  V1110 Tau                 & Binary\tablenotemark{f} & YSO        & 11.881$\pm$0.149 & -52.705$\pm$0.062 & -11.321$\pm$0.066 & 46 & 11.6$\pm$0.04    & -52.267$\pm$0.093  & -11.342$\pm$0.07    & 0.0    & 9.959  & 1.189\\
  55  &  V999 Tau                  &        & YSO        & 6.972$\pm$0.197  & 9.533$\pm$0.218   & -15.684$\pm$0.198 & 46 & 8.138$\pm$0.391  & 5.493$\pm$0.641    & -18.99$\pm$0.385    & 1.967  & 13.515 & 2.97\\
  56  &  V1000 Tau                 &    Binary\tablenotemark{g}      & YSO        & 7.324$\pm$0.132  & 6.01$\pm$0.235    & -17.72$\pm$0.159  & 46 &                  &                    &                     & 15.396 & 13.532 & 3.359\\
  57  & HD 282630                    &        & YSO        & 7.061$\pm$0.125  & 3.897$\pm$0.113   & -24.21$\pm$0.132  & 46 & 6.263$\pm$0.079  & 4.307$\pm$0.154    & -24.132$\pm$0.076   & 0.0    & 10.333 & 1.259\\
  58  & HD 283572                    &        & YSO        & 7.722$\pm$0.057  & 8.853$\pm$0.096   & -26.491$\pm$0.113 & 46 & 7.673$\pm$0.052  & 9.011$\pm$0.115    & -26.385$\pm$0.072   & 0.0    & 8.796  & 1.07\\
  58  &                              &        &            & 7.78$\pm$0.04    & 7.819$\pm$0.053   & -26.6$\pm$0.1     &  9 &                  &                    &                     &        &        & \\
  59  &  V1229 Ori                 &     Binary\tablenotemark{g}     & YSO        & 2.567$\pm$0.051  & 2.38$\pm$0.08     & 0.55$\pm$0.14     & 41 & 2.519$\pm$0.054  & 2.13$\pm$0.1       & 0.781$\pm$0.085     & 0.0    & 12.259 & 1.377\\
  60  &  MT Ori                    &        & YSO        & 2.646$\pm$0.041  & 3.82$\pm$0.1      & 1.6$\pm$0.17      & 41 & 2.404$\pm$0.053  & 3.772$\pm$0.096    & 2.345$\pm$0.08      & 0.0    & 11.324 & 1.61\\
  61  & * tet01 Ori E                & Binary\tablenotemark{f} & YSO        & 2.557$\pm$0.051  & 1.45$\pm$0.03     & 1.02$\pm$0.08     & 41 & 2.388$\pm$0.095  & 1.614$\pm$0.122    & 1.228$\pm$0.114     & 0.0    & 9.645  & 1.455\\
  62  & * tet01 Ori A                & Binary\tablenotemark{f} & YSO        & 2.626$\pm$0.1    & 4.81$\pm$0.1      & -2.53$\pm$0.12    & 41 & 2.373$\pm$0.105  & 1.536$\pm$0.158    & 0.123$\pm$0.139     & 0.227  & 6.609  & 0.151\\
  63  & Brun 656                     &     Binary\tablenotemark{g}     & YSO        & 2.708$\pm$0.21   & 2.36$\pm$0.69     & 0.06$\pm$1.05     & 41 & 2.114$\pm$0.046  & 0.111$\pm$0.073    & -3.978$\pm$0.08     & 0.0    & 10.483 & 1.489\\
  64  &  V1699 Ori                 &        & YSO        & 2.493$\pm$0.049  & 1.76$\pm$0.05     & -0.89$\pm$0.16    & 41 & 2.555$\pm$0.262  & 1.551$\pm$0.425    & -1.034$\pm$0.366    & 1.19   & 18.396 & 2.388\\
  65  &  V1961 Ori                 &        & YSO        & 2.533$\pm$0.027  & -7.22$\pm$0.06    & -0.99$\pm$0.08    & 41 & 2.546$\pm$0.041  & -7.194$\pm$0.089   & -1.037$\pm$0.09     & 0.0    & 11.609 & 1.61\\
  66  &  V1321 Ori                 &        & YSO        & 2.509$\pm$0.044  & 0.06$\pm$0.12     & 6.95$\pm$0.16     & 41 & 2.431$\pm$0.037  & 0.259$\pm$0.059    & 7.103$\pm$0.057     & 0.0    & 10.13  & 1.623\\
  67  & Brun 334                     & Binary\tablenotemark{f} & YSO        & 2.591$\pm$0.046  & -4.01$\pm$0.08    & -1.17$\pm$0.07    & 41 & 2.495$\pm$0.043  & -3.848$\pm$0.1     & -0.958$\pm$0.083    & 0.0    & 10.818 & 1.58\\
  68  & HD 37150                     &        & YSO        & 2.536$\pm$0.046  & 1.32$\pm$0.05     & -0.56$\pm$0.12    & 41 & 2.706$\pm$0.061  & 1.712$\pm$0.133    & -0.194$\pm$0.106    & 0.233  & 6.508  & -0.284\\
  69  &  V1046 Ori                 & Binary\tablenotemark{f} & YSO        & 2.643$\pm$0.075  & 1.88$\pm$0.09     & 1.2$\pm$0.14      & 41 & 0.447$\pm$0.171  & 0.459$\pm$0.394    & 2.511$\pm$0.384     & 0.62   & 6.532  & -0.199\\
  70  &  V363 Ori                  &        & YSO        & 2.575$\pm$0.389  & 0.33$\pm$0.05     & -1.34$\pm$0.43    & 41 & 3.181$\pm$0.182  & 2.606$\pm$0.343    & -0.431$\pm$0.31     & 1.361  & 13.574 & 2.552\\
  71  &  V1727 Ori                 &        & YSO        & 2.312$\pm$0.207  & 2.54$\pm$0.3      & -1.3$\pm$0.64     & 41 & 2.196$\pm$0.233  & 2.872$\pm$0.399    & -1.016$\pm$0.347    & 1.622  & 13.783 & 2.212\\
  72  &  V621 Ori                  &        & YSO        & 2.422$\pm$0.034  & 0.19$\pm$0.44     & -0.97$\pm$0.27    & 41 & 2.691$\pm$0.11   & 0.665$\pm$0.144    & -1.305$\pm$0.114    & 0.368  & 14.183 & 2.694\\
  73  & HD 294300                    &   Binary\tablenotemark{g}       & YSO        & 3.303$\pm$0.353  & -4.92$\pm$0.66    & 4.67$\pm$1.37     & 41 & 2.789$\pm$0.12   & 2.775$\pm$0.172    & -3.188$\pm$0.132    & 0.195  & 9.718  & 1.014\\
  74  & HD 290862                    &     Binary\tablenotemark{g}     & YSO        & 2.197$\pm$0.545  & 0.35$\pm$0.27     & 0.83$\pm$0.83     & 41 & 2.177$\pm$0.064  & -0.257$\pm$0.109   & -0.652$\pm$0.104    & 0.16   & 10.18  & 1.072\\
  75  & 2MASS J05420800-0812028      &        & YSO        & 2.315$\pm$0.072  &  0.13$\pm$0.25    & -1.05$\pm$0.18    & 41 & 1.589$\pm$0.925  & 0.466$\pm$1.955    & -4.241$\pm$2.303    &  3.01  & 20.354 & 4.26\\
  76  & GBS-VLA J054643.62+000528.3  &        & YSO        & 2.608$\pm$0.047  & -1.02$\pm$0.02    & -0.52$\pm$0.15    & 41 & 2.381$\pm$0.062  & -0.54$\pm$0.114    & -1.183$\pm$0.103    &  0.33  & 13.992 & 2.79\\
  77  & TYC 5346-538-1               &        & YSO        & 2.348$\pm$0.069  &  0.68$\pm$0.09    & -0.51$\pm$0.25    & 41 & 2.393$\pm$0.06   & 0.827$\pm$0.145    & -0.322$\pm$0.146    &  0     & 10.647 & 1.21\\
  78  & [SSC75] M 78  11             &        & YSO        & 2.547$\pm$0.034  &  0.01$\pm$0.01    & -0.49$\pm$0.08    & 41 & 2.494$\pm$0.041  & 0.29 $\pm$0.076    & -1.185$\pm$0.072    &  0     & 12.513 & 1.84\\
  79  & [BCB89] IRS 11               &     Binary\tablenotemark{g}     & YSO        & 1.865$\pm$0.105  & -0.43$\pm$0.16    &  1.03$\pm$0.42    & 41 & 2.613$\pm$0.307  & 0.649$\pm$0.407    & -1.415$\pm$0.407    &  0.96  & 18.093 & 3.40\\
  80  & [BCB89] IRS 15               &     Binary\tablenotemark{g}    & YSO        & 2.223$\pm$0.121  &  0.04$\pm$0.31    &  0.2 $\pm$0.43    & 41 & 2.637$\pm$0.152  & 0.029$\pm$0.221    & -0.646$\pm$0.209    &  0.73  & 16.110 & 4.02\\
  81  & 2MASS J05413786-0154323      &    Binary\tablenotemark{g}      & YSO        & 2.804$\pm$0.032  &  0.55$\pm$0.1     & -0.1 $\pm$0.15    & 41 & 2.780$\pm$0.238  & -1.11$\pm$0.449    & -0.754$\pm$0.386    &  0.96  & 17.259 & 2.77\\
  82  & EM* LkHA 101                 &   Binary\tablenotemark{g}       & YSO        & 1.87 $\pm$0.1    &  1.86 $\pm$0.04   & -5.7 $\pm$0.05    & 13 & 1.614$\pm$0.078  & 2.015$\pm$0.154    & -5.320$\pm$0.115    &  0.38  & 14.854 & 2.977\\
  83  &  V913 Per                  &        & YSO        & 3.119$\pm$0.104  & 2.458$\pm$0.047   & -7.272$\pm$0.133  & 48 & 3.708$\pm$0.262  & 5.039$\pm$0.482    & -7.111$\pm$0.281    & 1.095  & 14.696 & 2.914\\
  84  &  V918 Per                  &      Binary\tablenotemark{g}    & YSO        & 3.129$\pm$0.512  & 4.857$\pm$0.335   & -6.75$\pm$0.488   & 48 & 1.852$\pm$0.333  & -3.321$\pm$0.602   & -9.831$\pm$0.439    & 0.818  & 12.097 & 2.209\\
  85  & Cl* IC 348 LRL 11            &     Binary\tablenotemark{g}     & YSO        & 2.68$\pm$0.076   & 2.37$\pm$0.08     & -8.271$\pm$0.16   & 48 & 2.665$\pm$0.117  & 1.814$\pm$0.214    & -9.807$\pm$0.123    & 0.475  & 14.087 & 3.019\\
  86  & IRAS 20126+4104              &        & YSO        & 0.61$\pm$0.02    &                   &                   & 18 & 0.886$\pm$0.368  & -3.913$\pm$0.611   & -4.578$\pm$0.701    & 1.548  & 19.19  & 2.355\\
  86  &                              &        &            & 0.75$\pm$0.092   & -4.15$\pm$0.51    & -4.07$\pm$0.51    & 36 &                  &                    &                     &        &        & \\
  87  & HD 36705                     &     Binary\tablenotemark{g}    & YSO        & 66.75$\pm$0.45   & 48.927$\pm$2.805  & 137.5$\pm$7.5     &  2 & 65.32$\pm$0.144  & 29.15$\pm$0.251    & 164.421$\pm$0.299   & 0.85   & 6.674  & 1.1\\
  88  &  T Tau                     &      Binary\tablenotemark{g}    & YSO        & 6.723$\pm$0.046  & 6.79$\pm$0.432    & -11.131$\pm$0.444 & 46 & 6.929$\pm$0.058  & 11.356$\pm$0.122   & -14.837$\pm$0.1     & 0.12   & 9.627  & 1.653\\
  88  &                                  &        &            & 6.90$\pm$0.09  & 12.35$\pm$0.04    & -12.80$\pm$0.06 & 50 &                  &                    &                     &        &        & \\
  89  &  V404 Cyg                  & Binary\tablenotemark{f} & Other Star & 0.418$\pm$0.024  & -5.04$\pm$0.02    & -7.64$\pm$0.03    & 16 & 0.439$\pm$0.1    & -5.769$\pm$0.175   & -7.845$\pm$0.173    & 0.501  & 17.188 & 2.874\\
  90  & LS I +61 303                 & Binary\tablenotemark{f} & Other Star & 0.26$\pm$0.61    & 0.967$\pm$0.26    & -1.21$\pm$0.32    &  3 & 0.38$\pm$0.038   & -0.296$\pm$0.041   & -0.079$\pm$0.067    & 0.0    & 10.393 & 1.29\\
  91  & * bet Per                    & Binary\tablenotemark{f} & Other Star & 33.32$\pm$0.73   & 2.787$\pm$0.136   & -0.64$\pm$0.18    &  3 &                  &                    &                     & 3.025  & 4.34   & 2.378\\
  91  &                              &        &            & 34.7$\pm$0.6     & 2.7$\pm$0.07      & -0.8$\pm$0.09     & 19 &                  &                    &                     &        &        & \\
  92  &  UX Ari                    & Binary\tablenotemark{f} & Other Star & 19.37$\pm$0.39   & 41.229$\pm$0.184  & -104.01$\pm$0.2   &  3 & 19.813$\pm$0.228 & 46.317$\pm$0.493   & -101.899$\pm$0.359  & 0.895  & 6.329  & 1.193\\
  92  &                              &        &            & 19.9$\pm$        & 44.96$\pm$0.13    & -102.33$\pm$0.09  & 19 &                  &                    &                     &        &        & \\
  93  & HD 22468                     & Binary\tablenotemark{f} & Other Star & 33.88$\pm$0.47   & -31.588$\pm$0.33  & -161.69$\pm$0.31  &  3 & 33.753$\pm$0.087 & -32.894$\pm$0.131  & -161.772$\pm$0.118  & 0.195  & 5.6    & 1.216\\
  94  &  BH CVn                    & Binary\tablenotemark{f} & Other Star & 22.21$\pm$0.45   & 85.496$\pm$0.131  & -9.22$\pm$0.16    &  3 & 21.669$\pm$0.16  & 85.607$\pm$0.189   & -9.711$\pm$0.148    & 0.702  & 4.73   & 0.633\\
  95  & * sig CrB                    & Binary\tablenotemark{f} & Other Star & 43.93$\pm$0.1    & -267.05$\pm$0.037 & -86.66$\pm$0.05   &  3 & 44.135$\pm$0.064 & -268.325$\pm$0.096 & -86.925$\pm$0.146   & 0.356  & 5.407  & 0.794\\
  96  & HD 226868                    & Binary\tablenotemark{f} & Other Star & 0.73$\pm$0.3     & -3.787$\pm$0.172  & -6.25$\pm$0.21    &  3 & 0.422$\pm$0.032  & -3.882$\pm$0.048   & -6.171$\pm$0.054    & 0.0    & 8.523  & 1.271\\
  96  &                              &        &            & 0.539$\pm$0.033  & -3.78$\pm$0.06    & -6.4$\pm$0.12     & 20 &                  &                    &                     &        &        & \\
  97  & HD 199178                    &        & Other Star & 8.59$\pm$0.33    & 26.595$\pm$0.407  & -1.24$\pm$0.43    &  3 & 8.902$\pm$0.04   & 26.323$\pm$0.061   & -0.742$\pm$0.066    & 0.0    & 7.005  & 1.018\\
  98  &  AR Lac                    & Binary\tablenotemark{f} & Other Star & 23.97$\pm$0.37   & -52.08$\pm$0.126  & 47.03$\pm$0.19    &  3 & 23.433$\pm$0.03  & -52.19$\pm$0.043   & 47.19$\pm$0.046     & 0.0    & 5.894  & 0.958\\
  99  &  IM Peg                    & Binary\tablenotemark{f} & Other Star & 10.28$\pm$0.62   & -20.587$\pm$0.459 & -27.53$\pm$0.4    &  3 & 10.05$\pm$0.091  & -20.719$\pm$0.137  & -26.851$\pm$0.131   & 0.277  & 5.655  & 1.355\\
  99  &                              &        &            & 10.37$\pm$0.07   & -20.83$\pm$0.09   & -27.27$\pm$0.09   & 25 &                  &                    &                     &        &        & \\
  100 &  AM Her                    & Binary\tablenotemark{f} & Other Star & 11.29$\pm$0.08   & -46.02$\pm$0.22   & 28.83$\pm$0.18    & 47 & 11.395$\pm$0.018 & -45.957$\pm$0.034  & 28.046$\pm$0.034    & 0.0    & 13.58  & 1.01\\
  101 & TVLM 513-46                  &   Binary\tablenotemark{g}   & Other Star & 93.27$\pm$0.18   & -39.75$\pm$0.074  & -65.5$\pm$0.07    & 44 & 93.45$\pm$0.195  & -43.766$\pm$0.349  & -63.997$\pm$0.329   & 1.105  & 16.53  & 5.031\\
  101 &                              &        &            & 92.92$\pm$0.23   & -39.225$\pm$0.111 & -65.47$\pm$0.12   & 29 &                  &                    &                     &        &        & \\
  102 &  SS Cyg                    & Binary\tablenotemark{f} & Other Star & 8.8$\pm$0.12     & 112.42$\pm$0.07   & 33.38$\pm$0.07    & 31 & 8.724$\pm$0.049  & 112.373$\pm$0.113  & 33.589$\pm$0.094    & 0.0    & 11.689 & 1.163\\
  103 & PSR J0437-4715               & Binary\tablenotemark{f} & Pulsar     & 6.396$\pm$0.054  & 121.679$\pm$0.052 & -71.82$\pm$0.086  & 10 & 8.325$\pm$0.678  & 122.864$\pm$1.197  & -71.166$\pm$1.67    & 0.0    & 20.41  & 1.548\\
  104 & PSR J1023+0038               & Binary\tablenotemark{f} & Pulsar     & 0.731$\pm$0.022  & 4.76$\pm$0.03     & -17.34$\pm$0.04   & 24 & 0.728$\pm$0.143  & 4.751$\pm$0.135    & -17.348$\pm$0.135   & 0.0    & 16.265 & 0.776\\
 105 & PN K 3-35               &        & AGB         & 0.26$\pm$0.04	& -3.11$\pm$0.10 & -5.93$\pm$0.07   & 1 & 0.384$\pm$0.125	& -2.935$\pm$0.149 & -3.351$\pm$0.185 & 2.410	& 17.546 &  2.159\\
  106 & IRAS 20143+3634         &        & YSO         & 0.367$\pm$0.037 & -2.99$\pm$0.16 & -4.37$\pm$0.43  & 4 & 0.320$\pm$0.071	& -3.113$\pm$0.108 & -2.923$\pm$0.147 & 0.000	& 16.932 &  1.785\\
  107 & BD+40 4220              & Binary\tablenotemark{f} & Other Star  & 0.61$\pm$0.22	& -1.64$\pm$0.98 & -7.16$\pm$0.98   & 6 & 0.638$\pm$0.056	& -3.083$\pm$0.101 & -4.085$\pm$0.080 & 0.000	& 8.240	 &  2.336\\
  108 & SV Peg                 &        & AGB         & 3.00$\pm$0.06	& 11.59$\pm$0.54 & -8.63$\pm$0.44   & 15 & 1.124$\pm$0.283	& 15.244$\pm$0.487 & -5.989$\pm$0.498 & 628.467	& 7.846	 &     \\
     \enddata
\end{deluxetable*}
\tablenotetext{a}{
 References:
1 --\citet{2011PASJ...63...71T},
 2--\citet{1997ApJ...490..835G},
 3--\citet{1999A&A...344.1014L},
 4--\citet{2014PASJ...66..102B},
 5--\citet{2003A&A...407..213V},
 6--\citet{2013ApJ...763..139D},
 7--\citet{2005ApJ...627L..49K},
 8--\citet{2007A&A...472..547V},
 9--\citet{2007ApJ...671.1813T},
10--\citet{2008ApJ...685L..67D},
11--\citet{2008PASJ...60.1007C},
12--\citet{2008PASJ...60.1013N},
13--\citet{2018ApJ...853...99D},
14--\citet{2009ApJ...698..242T},
15--\citet{2019PASJ...71...16S},
16--\citet{2009ApJ...706L.230M},
17--\citet{2010ApJ...721..267A},
18--\citet{2011A&A...526A..66M},
19--\citet{2011ApJ...737..104P},
20--\citet{2011ApJ...742...83R},
21--\citet{2011PASJ...63...63N},
22--\citet{2012A&A...544A..42Z},
23--\citet{2012ApJ...744...23Z},
24--\citet{2012ApJ...756L..25D},
25--\citet{2012ApJS..201....5R},
26--\citet{2012PASJ...64....7K},
27--\citet{2012PASJ...64..142I},
28--\citet{2013ApJ...774..107K},
29--\citet{2013ApJ...777...70F},
30--\citet{2013PASJ...65...28I},
31--\citet{2013Sci...340..950M},
32--\citet{2014PASJ...66...38M},
33--\citet{2014PASJ...66..101N},
34--\citet{2014PASJ...66..107K},
35--\citet{2014Sci...345.1029M},
36--\citet{2015PASJ...67...66N},
37--\citet{2016PASJ...68...71K},
38--\citet{2016PASJ...68...75K},
39--\citet{2016PASJ...68...78N},
40--\citet{2017ApJ...834..141O},
41--\citet{2017ApJ...834..142K},
42--\citet{2017ApJ...834..143O},
43--\citet{2017ApJ...849...99Z},
44--\citet{2017MNRAS.466.4211G},
45--\citet{2018ApJ...859...14X},
46--\citet{2018ApJ...859...33G},
47--\citet{2018MNRAS.475.1399G},
48--\citet{2018arXiv180803499O},
49--\citet{2012ApJ...747...18T},
50--\citet{2007ApJ...671..546L}}
\tablenotetext{b}{$\mu_x=\mu_{\alpha} \cos{\delta}$, $\mu_y=\mu_{\delta}$}
\tablenotetext{c}{astrometric ``excess noise''}
\tablenotetext{d}{G-band mean magnitude}
\tablenotetext{e}{\Gaia\ color: $G_{BP}$-$G_{RP}$}
\tablenotetext{f}{The binary was identified by SIMBAD database.}
\tablenotetext{g}{The binary was identified by each reference paper. }
\end{longrotatetable}

\clearpage
\bibliographystyle{aasjournal}
\bibliography{ref}

\end{CJK*}
\end{document}